\documentstyle[aps,prl,preprint,psfig]{revtex}
\begin{document}
\title{Coefficient of normal restitution of viscous particles and
  cooling rate of granular gases} \author{Thomas
  Schwager\cite{bylinemail} and Thorsten P\"oschel\cite{bylinemail}}
\address{Humboldt-Universit\"at zu Berlin,
  Institut f\"ur Physik, \\
  Invalidenstra\ss e 110, D-10115 Berlin, Germany} \draft
\date{\today} \maketitle

\begin{abstract}
  We investigate the cooling rate of a gas of inelastically
  interacting particles. When we assume velocity dependent
  coefficients of restitution the material cools down slower than with
  constant restitution. This behavior might have large influence to
  clustering and structure formation processes.
\end{abstract}

The behavior of granular gases has been of large scientific interest
in recent time. Goldhirsch and Zanetti~\cite{Goldhirsch} and McNamara
and Young~\cite{McNamara} have shown that a homogeneous granular gas
is unstable. After some time one observes dense regions (clusters) and
voids. To evaluate the loss of mechanical energy due to collisions one
introduces the coefficient of (normal) restitution
\begin{equation}
  \label{restitution}
  g^\prime = \epsilon g~,
\end{equation}
where $g=\left|\vec{g}\right|$ and $g{'}=\left|\vec{g}^{'}\right|$,
describing the loss of relative normal velocity $g^\prime$ of a pair
of colliding particles after the collision with respect to the impact
velocity $g$

It can be shown that even for three particles for certain region of
the coefficient of restitution there exist initial conditions which
lead to a behavior which is called ``inelastic collapse''. This means
that the particles accomplish an infinite number of collisions in
finite time~\cite{McNamara}. The conditions under which one can
observe inelastic collapse have been studied in one dimensional
systems~\cite{CGM} as well as in higher
dimension~\cite{Kadanoff}. Recently it was shown numerically that the
probability for a collapse rises significantly when the particles have
rotational degree of freedom~\cite{Schoerghofer}. In this case the
collapse is possible for much larger coefficients of restitution than
for non rotating particles. Other interesting related results concern
bouncing ball experiments on vibrating tables where complicated
dynamical behavior is observed (e.g.~\cite{Mehta}). Recently
complicated, under certain circumstances irregular motion of a
bouncing cantilever of an atomic force microscope when excited by a
transducer was investigated~\cite{BKG}.

In the
investigations~\cite{Goldhirsch,McNamara,CGM,Kadanoff,Schoerghofer,Mehta,BKG}
the approximation of constant coefficient of restitution was assumed.
Solving viscoelastic equations for spheres currently it was shown that
the coefficient of normal restitution $\epsilon$ is not a constant but
a function of the impact velocity $\epsilon\left(g\right)$
itself\cite{BSHP,KK}. For the ``compression'' $\xi = R_1+R_2
-\left|\vec{r}_1 - \vec{r}_2\right| $ of particles with radii $R_1$
and $R_2$ at positions $\vec{r}_1$ and $\vec{r}_2$ one finds
\begin{equation}
\ddot \xi +\rho\left( \xi^{3/2} +\frac{3}{2}\,A\, 
\sqrt{\xi}\, \dot {\xi}\right)=0
\label{xitaylor}
\end{equation}
\begin{equation}
  \rho=\frac{2~ Y\sqrt{R^{\,\mbox{\it \footnotesize eff}}}}
{3~ m^{\mbox{\it \footnotesize eff}}\left( 1-\nu ^2\right) }
\end{equation}
 $Y$ is the Young modulus, $\nu$ the Poisson ratio and 
 \begin{mathletters}
   \begin{eqnarray}
     m_{\mbox{\it \footnotesize eff}}&=&\frac{m_1 m_2}{m_1+m_2}\\ 
     R_{\mbox{\it \footnotesize eff}}&=&\frac{R_1 R_2}{R_1+R_2} 
   \end{eqnarray}
\end{mathletters}
are the effective radius and mass of the grains, respectively. $A$ is
a material constant depending on the Young modulus, the viscous
constants and the Poisson ratio of the material. Equation
(\ref{xitaylor}) was derived under the precondition that the colliding
spheres have impact velocity much less than speed of sound in the
particle material. For details see \cite{BSHP}. The initial conditions
for solving (\ref{xitaylor}) are
\begin{mathletters}
  \begin{eqnarray}
    \xi(0) &=& 0 \\
    \dot{\xi}(0) &=& g~~.  
  \end{eqnarray}
\end{mathletters}
The coefficient of restitution $\epsilon$ of at time $t=0$ colliding
spherical grains can be found from this equation relating the relative
normal velocities $g = \dot \xi(0)$ at time of impact and at time
$t_c$, when the particles separate after the collision, i.e. $t_c$ is
the collision time:
\begin{equation}
  \label{eps1}
  \epsilon=-\dot{\xi} \left(t_c\right)/\dot{\xi}\left(0\right).
\end{equation}
The (numerical) integration of equation (\ref{eps1}) yields the
coefficient of restitution as a function of the impact velocity (see
fig.~1 in \cite{BSHP}) which is in good agreement with experimental
data\cite{Bridges}. Constant coefficient of restitution, however, does
{\em not} agree with experimental experience~\cite{Hatzes}. Other
theoretical work on this topic can be found e.g.
in~\cite{Johnson,pao}.

Consider a gas of granular particles at a given initial granular
Temperature $T_0$. Then the question arises how the temperature
decreases with time due to inelastic collisions. This problem has been
investigated earlier~\cite{Haff,LunSavage} for the case of constant
coefficient of restitution and the result is (s.
also~\cite{EsipovPoeschel:1996})
\begin{equation}
  \label{Tconst}
  T(t) = T_0\,\left(1+t/\tau\right)^{-2}~.
  \label{T}
\end{equation}
The time scale $\tau$ is a material constant. The temperature decay
(\ref{Tconst}) is the origin of the cluster instabilities which have
been investigated recently~\cite{Goldhirsch,McNamara}.

The aim of the present paper is to derive an explicite analytic
expression for the coefficient of normal restitution $\epsilon(g)$ as
a function of the impact velocity $g$. A direct consequence of this
result will be a refined expression for the temperature decay of a
granular gas.

\begin{figure}[htbp]
\centerline{\psfig{figure=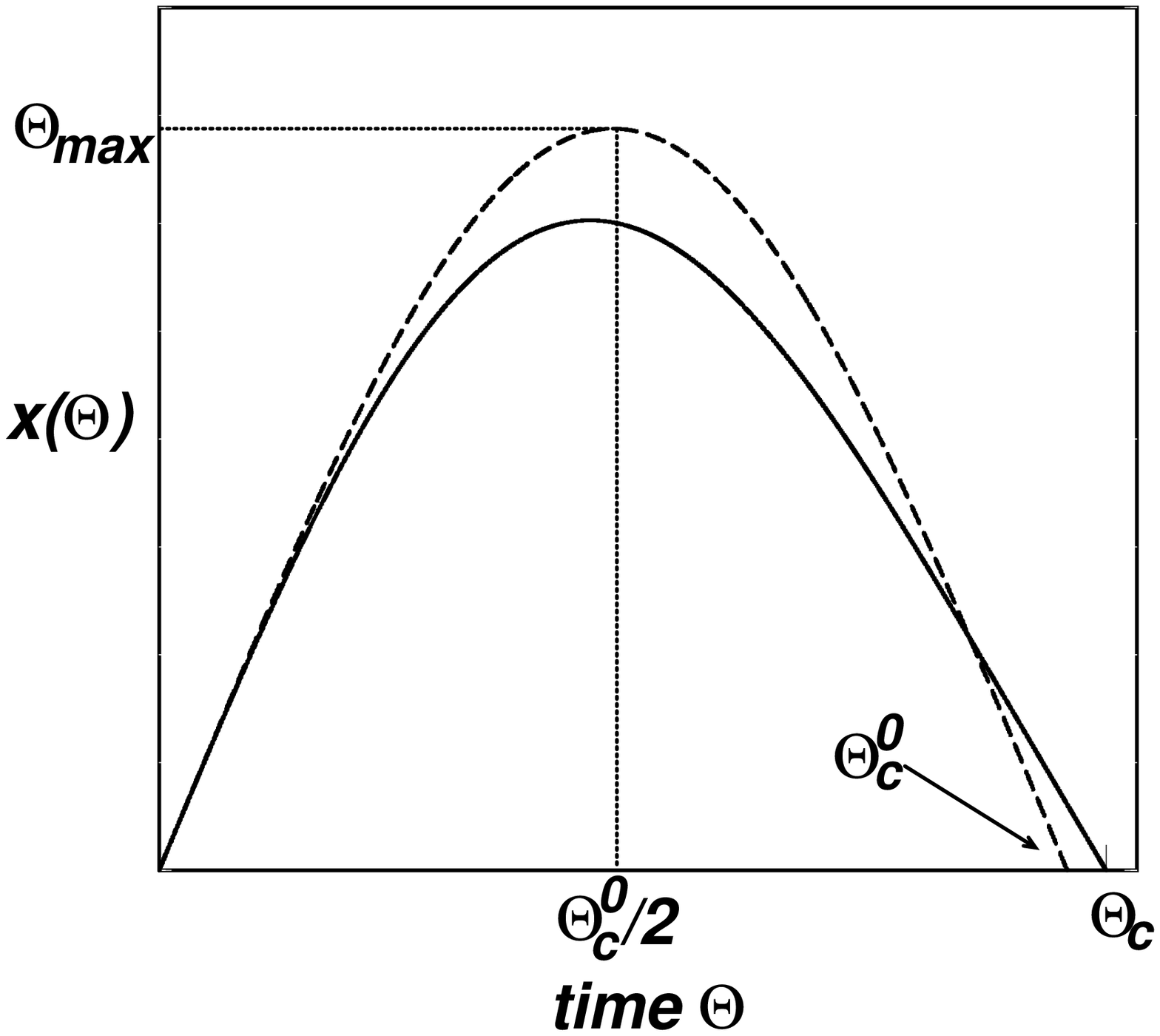,width=16cm,angle=0}}
\caption{The dynamics of the collision. The dashed line shows the 
  (strictly symmetric) solution of the undamped collision. For the
  case of the damped motion (full line) the maximum penetration depth
  is achieved earlier whereas the duration of the collision is longer
  ($\Theta_c>\Theta_c^0$).}
\label{fig:trajectory}
\end{figure}

The duration of collision $t_c^{0}$ for the undamped problem ($A=0$)
is given by~\cite{Hertz:1882}:
\begin{equation}
  \label{Thetadef}
  t_c^{0}=\frac{\Theta_c^{0}}{\rho^{\frac{2}{5}}g^{\frac{1}{5}}} ~.
\end{equation}
We want to point out here that $\Theta_c^0$ is a constant pure number,
{\em not} depending on any material properties. Hence, $t_c^0$ depends
only on the material constant $\rho$ and on the initial velocity $g$.
We use equation (\ref{Thetadef}) to define a rescaled dimensionless
time $\Theta$:
\begin{equation}
  \Theta=\rho^{\frac{2}{5}}~g^{\frac{1}{5}}~t
\end{equation}
Using the abbreviations
\begin{mathletters}
  \begin{eqnarray}
    v&=&\rho^{2}g\\
    \alpha&=&\frac{3}{2}A
  \end{eqnarray}
\end{mathletters}
and a new set of variables
\begin{mathletters}
  \begin{eqnarray}
    \Theta&=&\rho^{\frac{2}{5}}~g^{\frac{1}{5}}~t~=~v^{\frac{1}{5}}~t\\
    x(\Theta)&=&\rho^2\xi(t)
  \end{eqnarray}
\end{mathletters}
we rewrite (\ref{xitaylor}) in the form 
\begin{equation}
  \label{initial}
  \ddot{x} + \alpha v^{-\frac{1}{5}} \dot{x}\sqrt{x} + 
v^{-\frac{2}{5}}x^{\frac{3}{2}} = 0
\end{equation}
with $\dot{x}=\frac{d}{d\Theta}x$. We see that 
\begin{equation}
  \frac{dx}{dt}(0)=\frac{1}{\rho^2}\frac{d\xi}{dt}(0)=\frac{g}{\rho^2}=
v=v^{\frac{1}{5}}\frac{dx}{d\Theta}(0)
\end{equation}
Hence the initial conditions in our new variables $x$ and $\Theta$ read
\begin{mathletters}
  \begin{eqnarray}
    x(0) &=& 0\\
    \frac{dx}{d\Theta}(0) = \dot{x}(0) &=& v^{\frac{4}{5}}
  \end{eqnarray}
\end{mathletters}
Both equations of motion, (\ref{xitaylor}) and (\ref{initial}) become
special at $x=0$ or $\xi=0$, respectively, i.e. all derivatives of
third order and higher diverge. This will be shown for the case of
$x$:
\begin{eqnarray}
  \frac{d}{d\Theta}~\ddot{x} &=& - \frac{d}{d\Theta}\left( \alpha 
v^{-\frac{1}{5}} \dot{x}\sqrt{x} + v^{-\frac{2}{5}}x^{\frac{3}{2}}
\right)\nonumber\\
  &=&\alpha v^{-\frac{1}{5}}\left(\ddot{x}\sqrt{x}+\frac{\dot{x}}
{2\sqrt{x}}\right) - \frac{3}{2}v^{-\frac{2}{5}}\dot{x}\sqrt{x}
\end{eqnarray}
Hence
\begin{equation}
  \lim_{x\rightarrow 0}\frac{d^3}{d\Theta^3}x=\pm\infty~.
\end{equation}
and so are the higher derivatives. Because of this singularity we must
not expand $x$ in powers of $\Theta$. Because of the initial
conditions $x(\Theta)$ has the form
\begin{eqnarray}
  \label{xtoeta}
  x(\Theta)&=&v^{\frac{4}{5}}\Theta~(1+\eta(\Theta))\\
  \eta(0)&=&0~,
\end{eqnarray}
which defines the function $\eta(\Theta)$. Using transformation
(\ref{xtoeta}) we find
\begin{equation}
  \Theta\ddot{\eta} + 2 \dot{\eta} + \alpha v^{\frac{1}{5}} 
\Theta^{\frac{3}{2}}\dot{\eta}\sqrt{1+\eta} + \left(\alpha 
v^{\frac{1}{5}}\sqrt{\Theta} + \Theta^{\frac{3}{2}}\right) 
(1+\eta)^{\frac{3}{2}} = 0~.
  \label{etaeq}
\end{equation}
In (\ref{etaeq}) occur terms $\Theta^{0.5}$ and $\Theta^{1.5}$,
therefore we expand $\eta$ in powers of $\sqrt{\Theta}$
\begin{equation}
  \label{etaseries}
  \eta=\sum_{k=0}^\infty a_k~\Theta^{\frac{k}{2}}
\end{equation}
The first coefficient $a_0$ vanishes because of the initial condition
for $x$. When we require
\begin{equation}
  \dot{\eta}=\frac{a_1}{2\sqrt{\Theta}}+a_2+\dots
\end{equation}
to be finite at $\Theta=0$ the second coefficient $a_1$ must vanish as
well. With Taylor expansion of $\sqrt{1+\eta}$ and
$(1+\eta)^{\frac{3}{2}}$ for small $\eta$ we arrive at
\begin{equation}
  \label{firstTermseta}
  \eta = -\frac{4}{15} \alpha v^{\frac{1}{5}}  \Theta^\frac{3}{2} - 
  \frac{4}{35} \Theta^{\frac{5}{2}} + \frac{3}{70} \alpha v^{\frac{1}{5}} 
  \Theta^4 + \frac{1}{15} \alpha^2 v^{\frac{2}{5}} \Theta^3 \dots
\end{equation}
and therefore
\begin{eqnarray}
\label{firstTermsx}
  x &=& v^{\frac{4}{5}}~\Theta - \frac{4}{15}\alpha v~\Theta^{\frac{5}{2}} 
  - \frac{4}{35} v^{\frac{4}{5}} \Theta^{\frac{7}{2}} + \frac{1}{15} 
  \alpha^2 v^{\frac{6}{5}} \Theta^4\nonumber \\
  && + \frac{3}{70} \alpha v \Theta^5 - \frac{38}{2475} \alpha^3 
  v^{\frac{7}{5}} \Theta^{\frac{11}{2}} + \frac{1}{175} v^{\frac{4}{5}} 
  \Theta^6 + \dots
\end{eqnarray}
Rearranging the full series (\ref{firstTermsx}) one finds 
\begin{eqnarray}
  x &=& v^{\frac{4}{5}}\left(\Theta-\frac{4}{35}\Theta^{\frac{7}{2}}+ 
    \frac{1}{175}\Theta^6+\dots\right)\nonumber\\
  &+& \alpha v \left(-\frac{4}{15}\Theta^{\frac{5}{2}} + 
    \frac{3}{70}\Theta^5 + \dots\right)\nonumber \\
  &+& \alpha^2 v^{\frac{6}{5}}\left(\frac{1}{15} \Theta^4 + 
    \dots\right) + \dots\nonumber\\
  &=& v^{\frac{4}{5}}x_0(\Theta) + \alpha v x_1(\Theta) + \alpha^2 
  v^{\frac{6}{5}} x_2(\Theta) \dots \label{xkdef} 
\end{eqnarray}
$v^{\frac{4}{5}}x_0$ is the solution of the undamped (elastic)
collision (s. dashed line in fig.~\ref{fig:trajectory}).  The full
line in fig.~\ref{fig:trajectory} shows the damped motion according to
eq. (\ref{firstTermsx}). The direct numerical integration of eq.
(\ref{initial}) collapses with the full line.

For $x\left(\frac{1}{2}\Theta_c^0\right)$ where $\Theta_c^0$ is the
duration of the undamped collision one finds using (\ref{xkdef}):
\begin{eqnarray}
  x\left(\frac{\Theta_c^0}{2}\right) &=& v^{\frac{4}{5}}x_0
  \left(\frac{\Theta_c^0}{2}\right)+ \alpha v x_1
  \left(\frac{\Theta_c^0}{2}\right) + \alpha^2 
  v^{\frac{6}{5}}x_2\left(\frac{\Theta_c^0}{2}\right) + \dots\nonumber\\
  &=& v^{\frac{4}{5}} B_0 + \alpha v B_1 + \alpha^2 v^{\frac{6}{5}} 
  B_2 + \dots~~,
\end{eqnarray}
which we do not need now but later on. 

Note that the coefficients $B_k$ are constants, i.e. they do not
depend on $v$ nor on material constants.

Equations (\ref{xitaylor}) and (\ref{initial}), respectively, hold for
the entire collision. The collision starts with $v$ and ends with
$v^{'}$. For practical purposes we now define the term {\em inverse
  collision}. The inverse collision is a collision which starts at
time $\Theta_c$ with relative velocity $v^{'}$ and ends at time $0$
with relative velocity $v$, i.e. time runs in inverse direction during
the inverse collision. The equation of motion for $x^{inv}$, i.e. for
a collision in inverse time follows from (\ref{initial}). Since the
inverse collision starts with $v^{'}$ we have to replace $v$ by
$v^{'}$. Because of the time reversal we have to change the sign of
time derivatives of odd orders, i.e. $\dot{x}\rightarrow
-\dot{x}^{inv}$. The equation of motion for the inverse collision
reads
\begin{equation}
  \label{reversal}
  \ddot{x}^{inv} - \alpha~ (v^{'})^{-\frac{1}{5}} 
  \dot{x}^{inv}\sqrt{x^{inv}} + (v^{'})^{-\frac{2}{5}}
  \left(x^{inv}\right)^{\frac{3}{2}} = 0
\end{equation}
A motion due to eq. (\ref{reversal}) in normal time would be an
accelerated one. However, we shall mention here that eqs.
(\ref{initial}) and (\ref{reversal}) describe strictly the same
physical motion.  The solution $x^{inv}$ of the inverse problem can be
derived from the solution of the direct problem replacing
$\alpha\rightarrow -\alpha$ and $v\rightarrow v^{'}$.
\begin{equation}
  \label{xinv}
  x^{inv}(\Theta') = (v^\prime)^{\frac{4}{5}}x_0(\Theta') - 
  \alpha v^\prime x_1(\Theta') + \alpha^2 (v^\prime)^{\frac{6}{5}} 
  x_2(\Theta') \dots
\end{equation}
Now we determine the collision time $\Theta_c$ and the final velocity.
One direct method to calculate $\Theta_c$ would be to determine the
solution of $x(\Theta) = 0$ using Taylor expansion of $x$ in the
region close to $\Theta_c^0$. It can be seen easily that this method
fails since all derivatives of $\frac{d^n}{d\Theta^n}x$ with $n\ge 3$
diverge for $\Theta=\Theta_c^0$. Therefore $\Theta_c$ has to be
calculated by an indirect methode.

\begin{figure}[htbp]
  \centerline{\psfig{figure=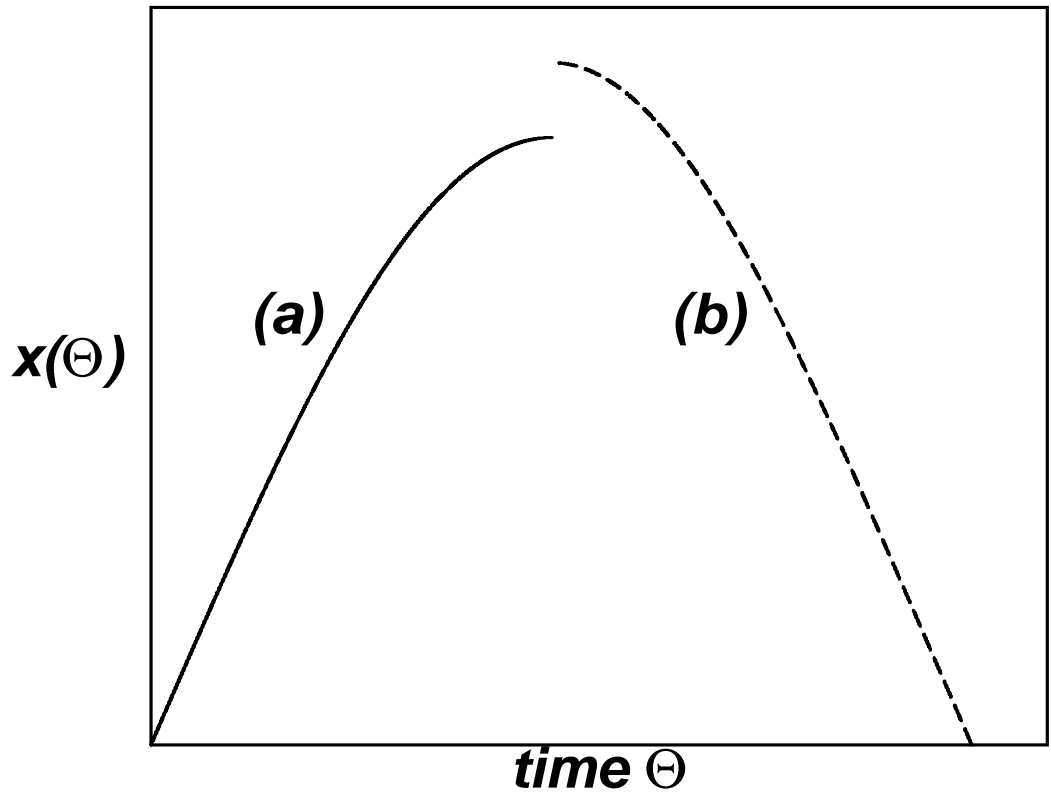,width=16cm,angle=0}}
\caption{Sketch of the calculation. The first part $(a)$ 
  $\Theta\in\left(0,\Theta_m\right)$ is calculated directly, for the
  other part $(b)$ we define the {\em inverse collision} where the
  particles start with velocity $v^\prime$ and velocity approaches
  zero at $\Theta=\Theta_m$. Both curves have to fit together
  smoothly.}
\label{fig:sketch}
\end{figure}

The problem will be subdivided into two parts (s.
fig.~\ref{fig:sketch}):
\begin{itemize}
\item[a)] the motion of the particles $x$ from $\Theta=0$ to time
  $\Theta_m$ when $x$ approaches its maximum and where $\dot{x}$
  changes its sign, and
\item[b)] from $\Theta_m$ to $\Theta_c$. 
\end{itemize}
In case of undamped motion where $\alpha=0$ we have $\Theta_m =
\Theta_c^0/2$. In part b) we do not consider the collision itself but
the inverse problem in the interval $(\Theta=0,~\Theta_m^\prime)$,
with $\Theta_m^\prime$ being the time where $x^{inv}$ approaches its
maximum. The continuity of both parts means $x\left(\Theta_m\right) =
x^{inv}\left(\Theta_m^\prime\right)$.

For finite damping $\alpha\ne 0$ we write $\Theta_m = \Theta_c^0/2 +
\delta$ and $\Theta^\prime_m = \left(\Theta_c^0\right)^\prime/2 +
\delta^\prime$ and remind that $\Theta_c^0 =
\left(\Theta_c^0\right)^\prime $ .  To get an expression for $\delta$
we expand
\begin{eqnarray}
  \dot{x}\left(\frac{\Theta_c^0}{2} + \delta \right) = 0 &=& 
\dot{x}\left(\frac{\Theta_c^0}{2}\right) + \delta \ddot{x}
\left(\frac{\Theta_c^0}{2}\right) + \frac{\delta^2}{2} 
\frac{d^3}{d\Theta^3} x\left(\frac{\Theta_c^0}{2}\right) + 
\dots \\
  &=& v^{\frac{4}{5}}\left(\dot{x}_0\left(\frac{\Theta_c^0}{2}\right) + 
\delta \ddot{x}_0\left(\frac{\Theta_c^0}{2}\right) + \frac{\delta^2}{2} 
\frac{d^3}{d\Theta^3} x_0\left(\frac{\Theta_c^0}{2}\right) + \dots\right) 
\nonumber\\
  &&+ v\alpha\left( \dot{x}_1\left(\frac{\Theta_c^0}{2}\right) + \delta 
\ddot{x}_1\left(\frac{\Theta_c^0}{2}\right) + \frac{\delta^2}{2} 
\frac{d^3}{d\Theta^3} x_1\left(\frac{\Theta_c^0}{2}\right)  + 
\dots\right)\nonumber \\
  &&+ v^{\frac{6}{5}}\alpha^2\left( \dot{x}_2\left(\frac{\Theta_c^0}{2}\right)
 + \delta \ddot{x}_2\left(\frac{\Theta_c^0}{2}\right) + \frac{\delta^2}{2} 
\frac{d^3}{d\Theta^3} x_2\left(\frac{\Theta_c^0}{2}\right) + \dots\right) 
\label{tmp1}
\end{eqnarray}
and using $\dot{x}_0\left(\Theta_c^0/2\right) = 0$
($v^\frac{4}{5}~x_0$ is the solution of the undamped problem)
\begin{equation}
  \delta = -\alpha v^{\frac{1}{5}} \frac{\dot{x}_1\left(\Theta_c^0/2\right)}
{\ddot{x}_0\left(\Theta_c^0/2\right)} + {\cal O} \left(\alpha^{2}\right)~.
  \label{delta}
\end{equation}
The expression (\ref{delta}) has to be inserted into the Taylor
expansion of $x\left(\Theta_c^0/2 + \delta \right)$:
\begin{eqnarray}
  x\left(\Theta_c^0/2 + \delta \right) 
  & =& v^{\frac{4}{5}}\left(x_0\left(\frac{\Theta_c^0}{2}\right) + \delta 
\dot{x}_0\left(\frac{\Theta_c^0}{2}\right) + \frac{\delta^2}{2} \ddot{x}_0
\left(\frac{\Theta_c^0}{2}\right) + \dots\right)\nonumber \\
  &&~~ + \alpha v \left(x_1\left(\frac{\Theta_c^0}{2}\right) + \delta 
\dot{x}_1\left(\frac{\Theta_c^0}{2}\right) + \frac{\delta^2}{2} 
\ddot{x}_1\left(\frac{\Theta_c^0}{2}\right)  + \dots\right)\\
  && = v^{\frac{4}{5}}x_0\left(\frac{\Theta_c^0}{2}\right) + \alpha 
v x_1\left(\frac{\Theta_c^0}{2}\right) - \frac{\alpha^2 v^{\frac{6}{5}}}
{2} \frac{\dot{x}_1^2\left(\Theta_c^0/2\right)}{\ddot{x}_0\left(
\Theta_c^0/2\right)} \nonumber\\
  &&~~ + \alpha^2 v^{\frac{6}{5}} x_2\left(\frac{\Theta_c^0}{2}\right) + 
{\cal O}\left(\alpha^3\right)~.
  \label{xmax1}
\end{eqnarray}
Hence
\begin{eqnarray}
  x\left(\Theta_m\right) = v^{\frac{4}{5}} x_0\left(\Theta_c^0/2\right) + 
\alpha v x_1\left(\Theta_c^0/2\right) + \alpha^2 v^{\frac{6}{5}} 
\left(x_2\left(\frac{\Theta_c^0}{2}\right)- \frac{1}{2}
\frac{\dot{x}_1^2\left(\Theta_c^0/2\right)}{\ddot{x}_0
\left(\Theta_c^0/2\right)}\right)+\dots 
\label{sol1}
\end{eqnarray}
Replacing again $v\rightarrow v^\prime$ and $\alpha\rightarrow
-\alpha$ yields
\begin{eqnarray}
  \delta^\prime &=& \alpha (v^\prime)^{\frac{1}{5}} 
\frac{\dot{x}_1\left(\Theta_c^0/2\right)}{\ddot{x}_0\left(\Theta_c^0/2\right)}
 + {\cal O} \left(\alpha^{2}\right)\\
  \label{deltas}
 x^{inv}\left(\Theta_m^\prime\right) &=& (v^\prime)^{\frac{4}{5}} 
x_0\left(\Theta_c^0/2\right) - \alpha v^\prime x_1\left(\Theta_c^0/2\right) +
 \alpha^2 (v^\prime)^{\frac{6}{5}} \left(x_2\left(\frac{\Theta_c^0}{2}\right)-
 \frac{1}{2}\frac{\dot{x}_1^2\left(\Theta_c^0/2\right)}{\ddot{x}_0
\left(\Theta_c^0/2\right)}\right)+\dots
  \label{sol2}
\end{eqnarray}
As explained above both solutions (\ref{sol1}) and (\ref{sol2}) have
to be equal. With
\begin{equation} 
  \beta = x_2\left(\frac{\Theta_c^0}{2}\right)- \frac{1}{2}
\frac{\dot{x}_1^2\left(\Theta_c^0/2\right)}{\ddot{x}_0
\left(\Theta_c^0/2\right)}
\end{equation}
we write
\begin{eqnarray}
v^{\frac{4}{5}} x_0\left(\frac{\Theta_0^c}{2}\right) + \alpha v x_1
\left(\frac{\Theta_0^c}{2}\right) + \alpha^2 v^{\frac{6}{5}}\beta = 
(v^\prime)^{\frac{4}{5}} x_0\left(\frac{\Theta_0^c}{2}\right) - 
\alpha v^\prime  x_1\left(\frac{\Theta_0^c}{2}\right) + \alpha^2 
(v^\prime)^{\frac{6}{5}}\beta ~. 
\end{eqnarray}
We expand $v^\prime$ in $\alpha$ 
\begin{equation}
  v^\prime = v + \alpha v_1 + \alpha ^2 v_2 + \dots ~,
\end{equation}
and find
\begin{eqnarray}
&&  v^{\frac{4}{5}} x_0\left(\frac{\Theta_0^c}{2}\right) + \alpha v 
x_1\left(\frac{\Theta_0^c}{2}\right) + \alpha^2 v^{\frac{6}{5}}\beta 
\nonumber\\
&=& v^{\frac{4}{5}}\left(1+ \frac{\delta v}{v}\right)^{\frac{4}{5}} 
x_0\left(\frac{\Theta_0^c}{2}\right) - \alpha v\left(1 + \frac{\delta v}
{v}\right) x_1\left(\frac{\Theta_0^c}{2}\right) + \alpha^2 v^{\frac{6}{5}}
\left(1 + \frac{\delta v}{v}\right)^{\frac{6}{5}} \beta 
\end{eqnarray}
with $\delta v = \alpha v_1 + \alpha^2 v_2 + \dots$. Writing
$(1+\frac{\delta v}{v})^{\frac{n}{5}}$ in powers of $\alpha$ and
comparing coefficients yields finally
\begin{eqnarray}
  \label{solvs}
  v^\prime &=& v \left(1+\frac{5}{2}\alpha v^{\frac{1}{5}}
\frac{x_1\left(\frac{\Theta_c^0}{2}\right)}{x_0\left(\frac{\Theta_c^0}{2}
\right)} + \frac{15}{4}\alpha^2 v^{\frac{2}{5}}
\left(\frac{x_1\left(\frac{\Theta_c^0}{2}\right)}{x_0
\left(\frac{\Theta_c^0}{2}\right)}\right)^2 + \dots \right)\nonumber \\
  &=&  v \left(1-\alpha v^{\frac{1}{5}} C_1 + \alpha^2 
v^{\frac{2}{5}} C_2 + \dots\right)~,
\end{eqnarray}
with 
\begin{mathletters}
  \begin{eqnarray}
    C_1&=&\frac{5}{2}\frac{x_1\left(\frac{\Theta_c^0}{2}\right)}
{x_0\left(\frac{\Theta_c^0}{2}\right)}\\
    C_2&=&\frac{15}{4}\left(\frac{x_1\left(\frac{\Theta_c^0}{2}\right)}
{x_0\left(\frac{\Theta_c^0}{2}\right)}\right)^2~. 
  \end{eqnarray}
  \label{Cdef}
\end{mathletters}
Since $\Theta_c^0$ does depend neither on any material properties nor
on the impact velocity $g$ or $v$, respectively, $C_1$ and $C_2$ are
pure numerical constants. Evaluating $C_1$ and $C_2$ in (\ref{Cdef})
numerically yields $C_1=1.15344$ and $C_2=0.79826$.

\begin{figure}[htbp]
    \centerline{\psfig{figure=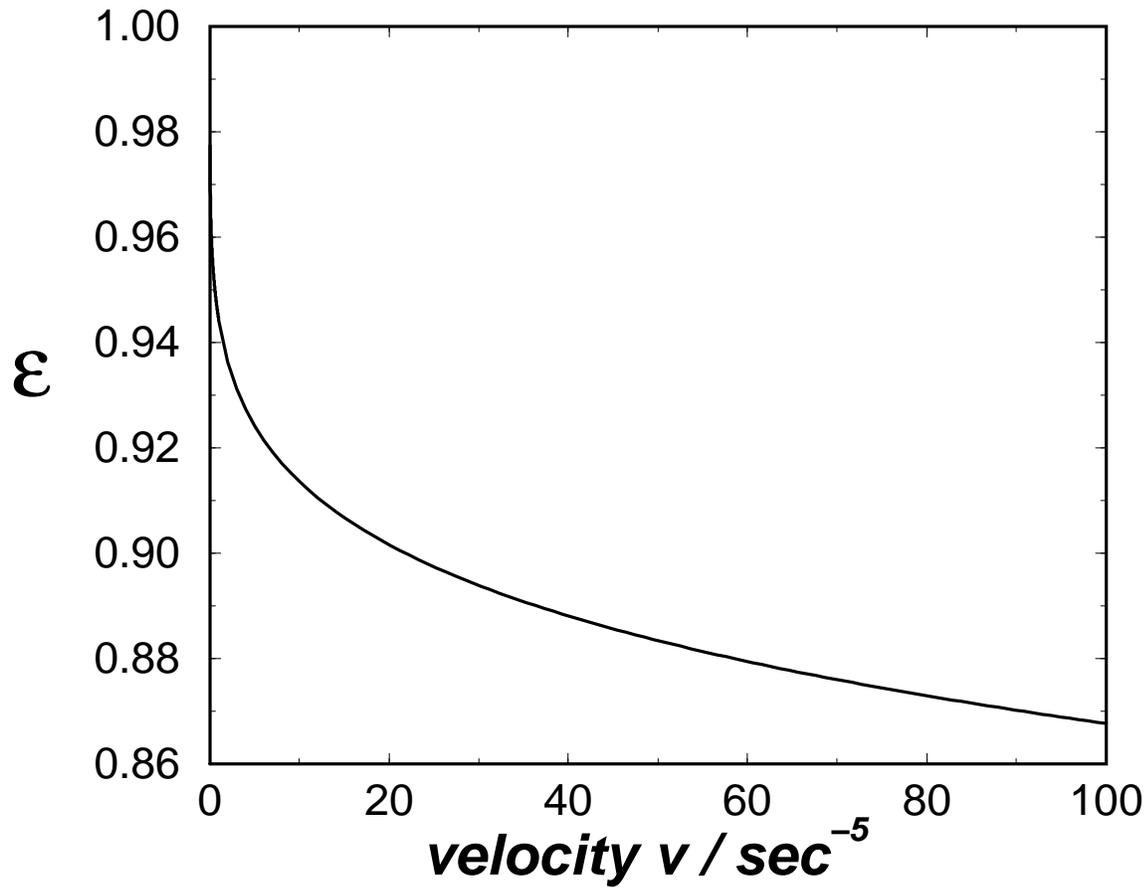,width=16cm,angle=0}}
\caption{The coefficient of restitution over impact velocity due 
  to eq. (\ref{epsfinal}). As expected for small relative velocity the
  particles collide almost elastically. The result of numerical
  integration of (\ref{eps1}) coincides with the curve. Both curves
  cannot be distinguished in the plot.}
\label{fig:ev}
\end{figure}

For coefficient of normal restitution one gets
\begin{mathletters}
  \begin{eqnarray}
    \epsilon = \frac{v^\prime}{v} &=& 1-\alpha v^{\frac{1}{5}} C_1 + 
\alpha^2 v^{\frac{2}{5}} C_2 + \dots\\
    &=& 1-C_1 A\rho^{\frac{2}{5}}g^{\frac{1}{5}} + 
    C_2 A^2\rho^{\frac{4}{5}}g^{\frac{2}{5}} + \dots 
  \end{eqnarray}
  \label{epsfinal}
\end{mathletters}
with $g$ being the impact velocity~(fig~\ref{fig:ev}). For the duration 
of collision we find with (\ref{delta}), (\ref{deltas}) and (\ref{solvs})
\begin{eqnarray}
  t_c&=&\left(\frac{\Theta_c^0}{2} + \delta\right) v^{-\frac{1}{5}} + 
\left(\frac{\Theta_c^0}{2} + \delta^\prime\right) 
(v^\prime)^{-\frac{1}{5}}\nonumber\\
  &=&\Theta_c^0v^{-\frac{1}{5}}\left(1 - \frac{1}{4}\alpha 
    v^{\frac{1}{5}}\frac{x_1\left(\Theta_c^0/2\right)}
    {x_0\left(\Theta_c^0/2\right)}\right) + {\cal O}(\alpha^2)\nonumber \\
  &=&\Theta_c^0v^{-\frac{1}{5}}\left(1+\frac{1}{10}C_1\alpha v^{\frac{1}{5}}
  \right) + {\cal O}(\alpha^2) \nonumber \\
  &=& \Theta_c^0 \rho^{-\frac{2}{5}}g~^{-\frac{1}{5}}\left(1+
    \frac{1}{10}C_1\alpha \rho^{\frac{2}{5}}g^{\frac{1}{5}}\right) + 
  {\cal O}(\alpha^2)~.\\
  \Theta_c &=& v^{\frac{1}{5}}t_c\nonumber\\
  &=&\Theta_c^0\left(1+\frac{1}{10}C_1\alpha v^{\frac{1}{5}}\right) + 
  {\cal O}(\alpha^2) 
  \label{Thetac}
\end{eqnarray}

To check the theoretical result (eqs. \ref{Cdef}) we integrated
numerically eq. (\ref{initial}) and received the curves $\epsilon(v)$
and $\Theta_c(v)$. Then we fitted $C_1$ and $C_2$ to these data using
(\ref{epsfinal}) and (\ref{Thetac}). For instance for $\alpha=0.05$ we
found $C_1^{\mbox{\footnotesize \em num}} =1.15356$ and
$C_2^{\mbox{\footnotesize\em num}}=0.80439$ from the curve
$\epsilon(v)$ (s. eq. \ref{epsfinal}). The fit of $C_1$ to
$\Theta_c(v)$ (s. eq. \ref{Thetac}) gives $C_1^{\mbox{\footnotesize\em
    num}}=1.15342$. For other values of $\alpha$ we found very similar
numbers. Hence, the numerical results agree with theory.

When we use the velocity dependent coefficient of restitution in the
collision term of Boltzmann equation
\begin{equation}
  \dot{T} \sim \int\int dv_1 dv_2 \left(1-\epsilon^2\right) \left| v_1 - 
    v_2 \right|^3 f\left(v_1\right) f\left(v_2\right)
\end{equation}
we get the cooling rate for dissipative gas
\begin{equation}
  T\sim T_0/\left(1+\frac{t}{\tau^\prime}\right)^{\frac{5}{3}}
  \label{cool}
\end{equation}
Our final results eq.~(\ref{epsfinal}) shows that for viscoelastic
colliding smooth bodies the coefficient of normal restitution is a
decreasing function with rising impact velocity: $1-\epsilon \sim
g^\frac{1}{5}$. A direct consequence is the cooling rate of a granular
gas (eq.~(\ref{cool})): a granular gas consisting of viscoelastic
particles cools down significantly {\em slower} than a gas of
particles which collide with constant coefficient of restitution
(s.~eq.~(\ref{T})). Due to our understanding it is neither
self-evident whether the clustering observed in granular gases of the
latter type, and the extreme case of this effect, the inelastic
collapse, will change their overall behavior nor whether they exist at
all. These questions should be reconsidered in detail for velocity
dependent restitution.

The authors are grateful to N.~Brilliantov, S.~Esipov, H.~Herrmann,
F.~Spahn and W.~Young for helpful discussions, and J.-M. Hertzsch for
providing relevant literature.

\end{document}